\input{aipcheck}

\documentclass[
    ,final            % use final for the camera ready runs
  ]
  {aipproc}

\layoutstyle{6x9}

\begin{document}

\title{Probing BH mass and accretion through X-ray variabiliy in the CDFS}

\author{V. Allevato}{
  address={Max-Planck-Institut f\"{u}r Plasmaphysik, Boltzmannstrasse 2, D-85748 Garching, Germany}
  ,altaddress={Dept. of Physical Sciences, University Federico II, via Cinthia 6, 80126 Naples, Italy}
}
\author{C. Pinto}{
  address={SRON, Sorbonnelaan 2, 3584 CA Utrecht, the Netherlands}
  ,altaddress={Dept. of Physical Sciences, University Federico II, via Cinthia 6, 80126 Naples, Italy}
}
\author{M. Paolillo}{
  address={Dept. of Physical Sciences, University Federico II, via Cinthia 6, 80126 Naples, Italy}
  ,altaddress={INFN - Naples Unit, Dept. of Physical Sciences, via Cinthia 9, 80126, Naples, Italy}
}
\author{I. Papadakis}{
  address={University of Crete Dept Physics, P.O. Box 2208, GR 710 03 Heraklion, Greece}
}
\author{P. Ranalli}{
  address={INAF - Osservatorio Astronomico di Bologna}
} 
\author{A. Comastri}{
  address={INAF - Osservatorio Astronomico di Bologna}
}

\author{K. Iwasawa}{
  address={INAF - Osservatorio Astronomico di Bologna}
} 
\begin{abstract}
 Recent work on nearby AGNs has shown that X-ray variability is correlated with the mass and accretion rate onto the central SMBH. Here we present the application of the variability-luminosity relation to high redshift AGNs in the \emph{CDFS}, making use of \textit{XMM-Newton} observations. We use Monte Carlo simulations in order to properly account for bias and uncertainties introduced by the sparse sampling and the very low statistics. Our preliminary results indicate that BH masses span over the range $10^{5} \div 10^{9} M_{\odot}$ while accretion rates range from $10^{-3}$ up to $>1~\dot{m}_{Edd}$.
%If confirmed, these results may help in planning the observing strategy for future X-ray AGN surveys.

\end{abstract}

\maketitle

%%%%%%%%%%%%%%%%%%%%%%%%%%%%%%%%%%%%%%%%%%%%
%% MAINMATTER
%%%%%%%%%%%%%%%%%%%%%%%%%%%%%%%%%%%%%%%%%%%
\subsection{Measuring variability for sparsely sampled/low statistics AGNs}

Recent studies of the X-ray variability of nearby AGNs have shown that the Power Density Spectrum (PDS) presents a characteristic timescale that correlates with BH mass and accretion rate \cite{2003ApJ...593...96M,2002MNRAS.332..231U,2004NuPhS.132..122M}.
The extension of these results to distant AGNs is difficult due to the sparse sampling and the low
statistics. In such cases the \emph{Excess Variance} is commonly used to estimate the intrinsic lightcurve variance \cite{1997ApJ...476...70N};
however it represents a maximum likelihood variability estimator only for identical/normal distributed
measurements errors and uniform sampling. Thus it can be used as an alternative approach provided that the effects of non-optimal observing conditions are properly taken into account.
Here we concentrate on the first set of 8 CDFS XMM-Newton observations spanning from July 2001 to Jan. 2002 for a total exposure time of about 370 ksec, where we detect variability in 59 of the 170 sources.

We performed Monte Carlo simulations of AGNs lightcurves in order to quantify the bias of the variability estimator, modifying the original Timmer \& Koenig algorithm  \cite{1995A&A...300..707T} that generates red-noise data with a power-law density spectrum, reproducing the sampling pattern and uncertainties of the real XMM-Newton observation. Fig.1 (left panel) presents the excess variance distribution for 5000 simulations of the sparsely sampled lightcurve compared to the input value, fixed at 0.04 (i.e. 20\% rms). The measured excess variance underestimates the intrinsic variance and the uneven sampling results in large uncertainties ($\sim 100\%$), mainly because
the sparse sampling doesn't allow to measure the intrinsic mean count rate.We used these results to correct the bias introduced by the sampling pattern, rescaling the excess variance by a factor given by the ratio between the input and median output value derived from the simulations.

\begin{figure}
\centering
  \includegraphics[height=.247\textheight]{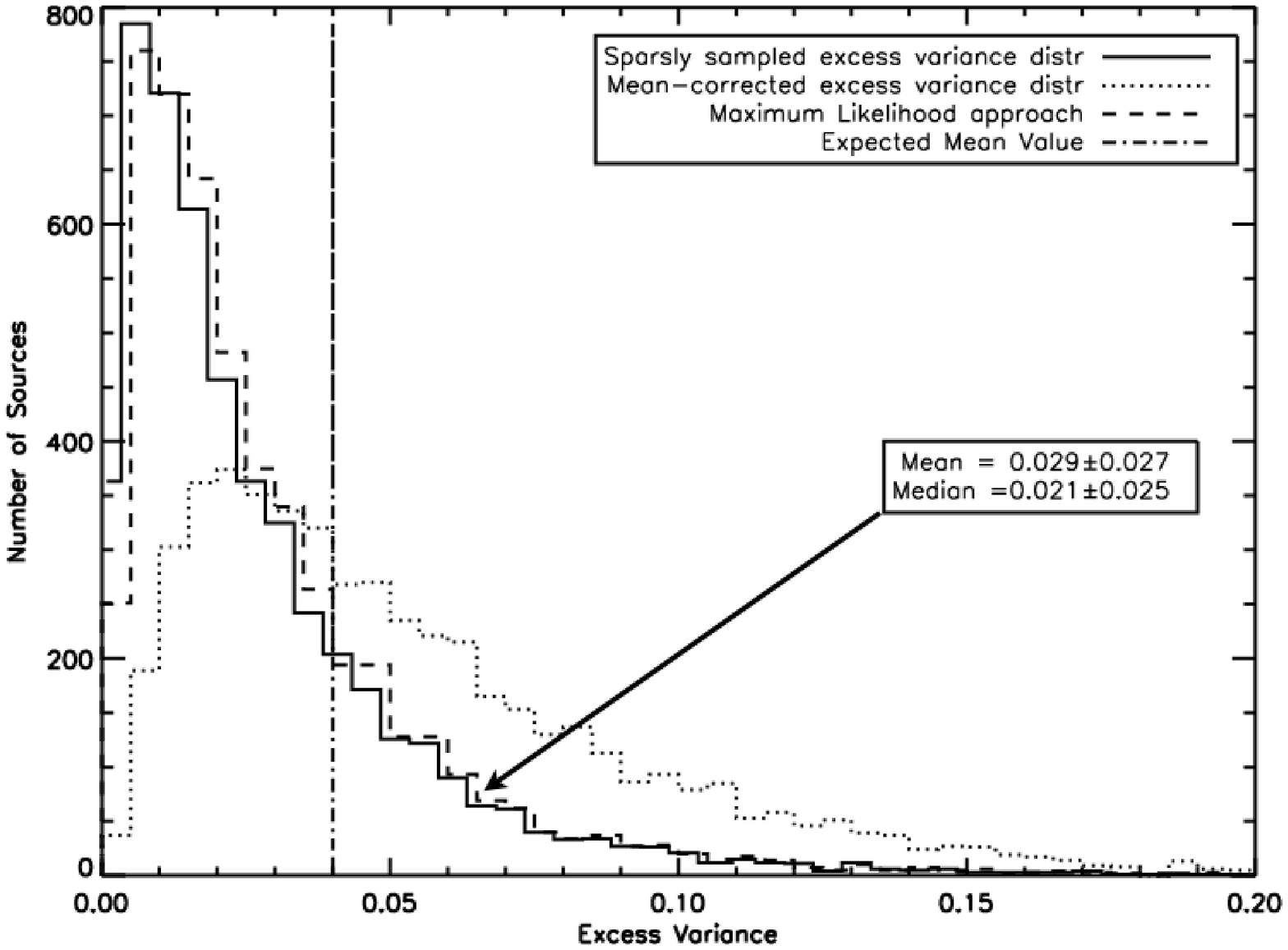}
  \includegraphics[height=.250\textheight]{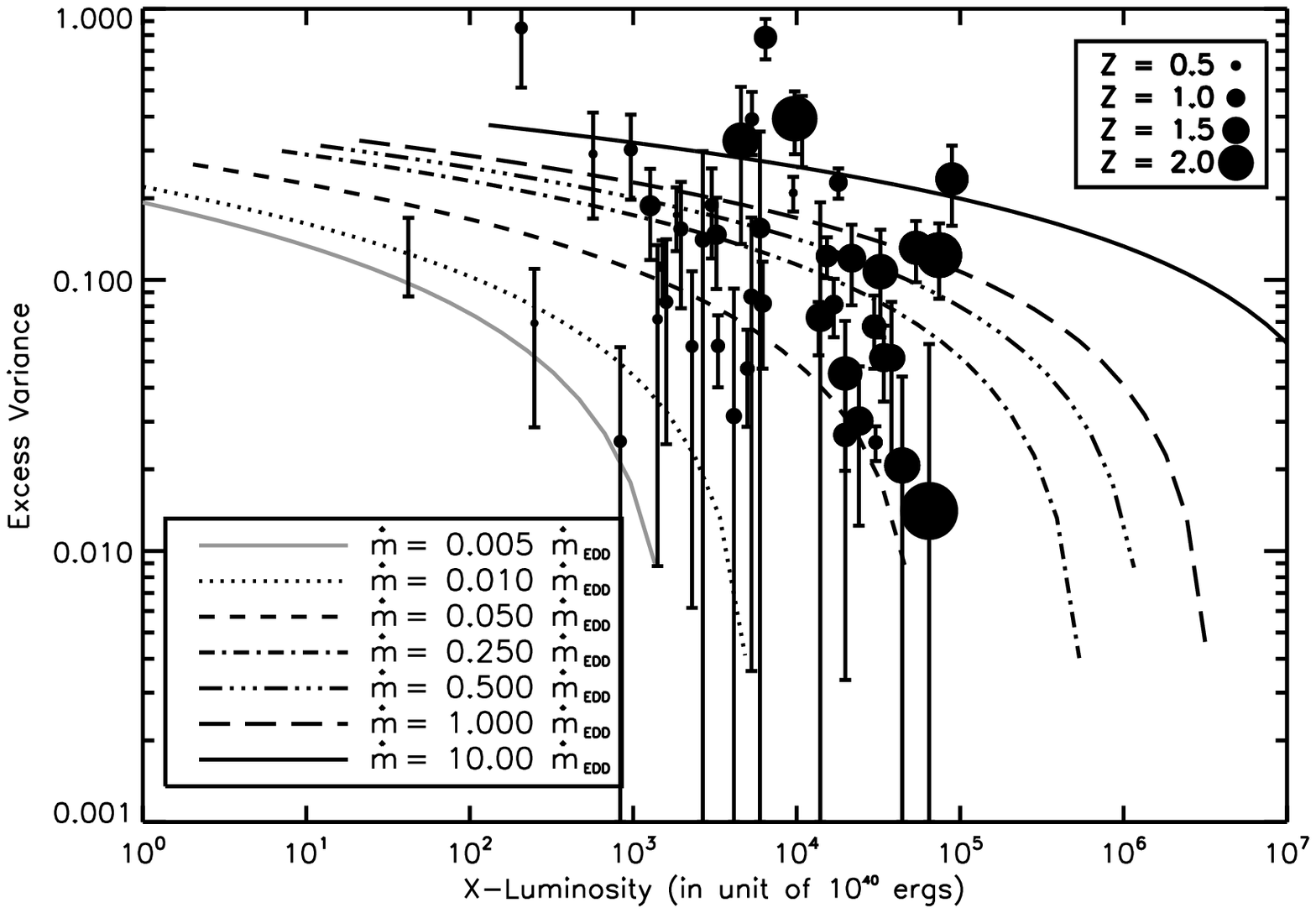}
  \caption{{\small \emph{Left}: Excess Variance distribution from 5000 simulations reproducing the CDFS sampling pattern (solid line), compared with the expected input value (vertical line), the distribution corrected for the mean count rate (dotted line), and the max.likelihood approach \cite{2000MNRAS.315..325A} (dashed line).
\emph{Right}: Luminosity-variability plot for the bright CDFS sources. The lines show the predictions of the model for different accretion rates values and a range of BH masses.\vspace{-0.5cm}}}
\end{figure}

\vspace{-0.5cm}\subsection{Mass and Accretion Rate Estimates}

We adopt the model by \cite{2008A&A...487..475P} to convert excess variance and X-ray luminosity into $\dot{m}_{Edd}$ and $M_{BH}$, assuming that the AGN PDSs is a power-law of slope -2 (-1) above (below) the break frequency:
$\sigma^{2}_{NXV} = \int_{lf}^{\infty} PDS(\nu) \mbox{d}\nu \propto \left[ ln(\nu_{bf}) - ln(\nu_{lf})+1 \right]$
where $\nu_{lf} = 1/T$ ($T$ is the length of the lightcurve) and $\nu_{bf}\propto\epsilon\dot{m}_{Edd} 10^6 M_{\odot}/M_{BH}$.
We can further relate the X-ray luminosity to the accretion rate and BH mass through  
%using $L_{bol}=1.5 \eta \dot{m}_{Edd} 10^{39} M_{BH}/M_{\odot}$ erg/s 
an appropriate X-ray to bolometric luminosity conversion.
% by, e.g., Marconi et al. 2006.
Fig.1 (right panel) shows the excess variance versus X-ray luminosity of our AGN sample, compared to the model predictions for different accretion rates values, and a range of BH masses. We obtain BH masses spanning over the range $10^{5} \div 10^{9} M_{\odot}$ and accretion rates from $10^{-3}$ up to $>1~\dot{m}_{Edd}$.

Our preliminary results suggest that X-ray variability can be used as a tool to measure
masses and accretion rates of AGNs in deep surveys, provided that the bias
and uncertainties introduced by sparse sampling and low statistics are properly accounted for.
We plan to extend our analysis to the additional 2.5 Msec CDF-S dataset recently
approved with XMM-Newton (PI A. Comastri).

\bibliographystyle{aipproc}   % if natbib is available

%%%Manual Bibliography. Comment if you want to use your bibtex database

%%%%%%%%%%%%%%%%%%%%%%%%%%%%%%%%%%%%%%%%%%%%
%%% You probably want to use your own bibtex database here
%%%%%%%%%%%%%%%%%%%%%%%%%%%%%%%%%%%%%%%%%%%%
%%Un comment If you want to use your database
%\bibliography{sample}
%%%%%%%%%%%%%%%%%%%%%%%%%%%%%%%%%%%%%%%%%%%%
%%% Just a reminder that you may have to run bibtex
%%% All of it up to \end{document} can be removed
%%% if you don't like the warning.
%  \typeout{******************************************}
%  \typeout{** Please run "bibtex \jobname" to optain}
%  \typeout{** the bibliography and then re-run LaTeX}
%  \typeout{** twice to fix the references!}
%  \typeout{******************************************}

\end{document}